\documentclass[sn-mathphys,Numbered]{sn-jnl}

\usepackage{graphicx}%
\usepackage{multirow}%
\usepackage{amsmath,amssymb,amsfonts}%
\usepackage{amsthm}%
\usepackage{mathrsfs}%
\usepackage[title]{appendix}%
\usepackage{xcolor}%
\usepackage{textcomp}%
\usepackage{manyfoot}%
\usepackage{booktabs}%
\usepackage{algorithm}%
\usepackage{algorithmicx}%
\usepackage{algpseudocode}%
\usepackage{listings}%

\theoremstyle{thmstyleone}%
\theoremstyle{thmstyletwo}%

\theoremstyle{thmstylethree}%

\begin{document}

\title[Article Title]{$^{27}$AL-NMR STUDY ON A SQUARE-KAGOME LATTICE ANTIFERROMAGNET}

\author*[1]{\fnm{Takayuki} \sur{Goto}}\email{gotoo-t@sophia.ac.jp}

\author[2]{\fnm{Masayoshi} \sur{Fujihala}}

\author[3]{\fnm{Setsuo} \sur{Mitsuda}}

\affil*[1]{\orgdiv{Faculty of Science and Technology}, \orgname{Sophia University}, \orgaddress{\street{Kioi-cho}, \city{Chiyoda-ku}, \postcode{102-8554}, \state{Tokyo}, \country{Japan}}}

\affil[2]{\orgdiv{Advanced Science Research Center}, \orgname{Japan Atomic Energy Agency (JAEA)}, \orgaddress{\street{Tokai-mura}, \city{Ibaraki}, \postcode{319-1195}, \country{Japan}}}

\affil[3]{\orgdiv{Department of Physics}, \orgname{Tokyo University of Science}, \orgaddress{\city{Shinjuku}, \postcode{162-8601}, \state{Tokyo}, \country{Japan}}}


\abstract{
NMR study has been performed on an $S=1/2$ antiferromagnet KCu$_6$AlBiO$_4$(SO$_4$)$_5$Cl 
on the square-Kagom\'{e} lattice, 
which has three slightly inequivalent nearest-neighbor interactions.  
Because of the geometrical frustration inherited from triangles within the square kagom\'{e} lattice
and of the low dimensionality, a long range magnetic order is strongly suppressed;
its absence has so far been confirmed in low temperatures down to dilution refrigerator region. 

$^{27}$Al-NMR spectra and the longitudinal relaxation time $T_1$ were measured by 
a conventional pulsed spectrometer on powder sample under 
several magnetic fields between 3 and 10 T and in low temperatures down to 0.35 K.  
The NMR line width due to the inhomogeneous broadening increased with lowering temperatures and leveled off below 3 K, 
where FWHM reached the value as large as 0.1 T, implying that the ground state is magnetic one, consistent with 
previous reports.
On the other hand, the longitudinal nuclear spin relaxation rate
$1/T_1$ obeyed the Arrhenius law with the thermal activation energy $\Delta =$ 2K at low temperatures,
suggesting that a small gap is formed in the spin excitation spectrum.
}




\maketitle

\section{Introduction}\label{sec1}

\begin{figure}[h]%
\centering
\includegraphics[width=4cm,bb=90 330 530 770,clip]{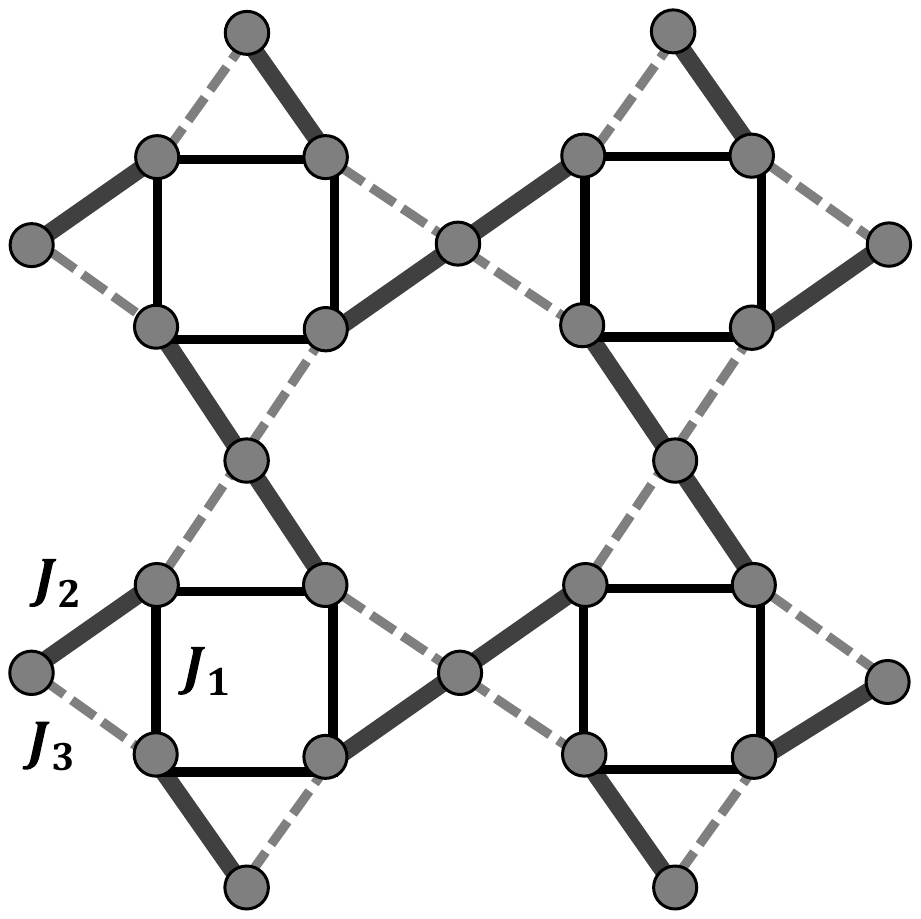}
\caption{Square-Kagom\'{e} lattice consisting of Cu$^2+$ ions with nearest-neighbor exchange couplings $J_1, J_2$ and $J_3$.
\cite{fujihala2020gapless,morita2018magnetization}}\label{fig1}
\end{figure}

The effect of geometrical frustration and low dimensionality of spin systems often brings us a possibility 
to find a new exotic state, 
because these two effects suppress the long range magnetic order 
and hence the paramagnetic temperature range is expected to be 
widen before entering the ordered state.
The title compound ${\rm KCu_6AlBiO_4(SO_4)_5Cl}$ consists of $S = 1/2 $ square-Kagom\'{e} lattice, possessing
a strong geometrical frustration effect, and is a good quasi-two dimensional system with
a long inter-plane distance of 10.27 ${\rm \AA}$\cite{fujihala2020gapless}.
So far, intensive studies have been performed on this compound both theoretically\cite{nakano2013_spin_flop,
Hasegawa_Sakai2018magnetization_jump,morita2018magnetization,Richter2022_calc,Pohle2023_CW} 
and experimentally\cite{fujihala2020gapless,Liu_specific_heat}.
It has been reported that the expected ground state depends delicately on the ratio of 
nearest-neighbor exchange couplings $J_1, J_2$ and $J_3$ (Fig. \ref{fig1})\cite{morita2018magnetization}, and
that no long range magnetic order takes place even at the dilution temperature range\cite{fujihala2020gapless}.
The large value of low-temperature uniform susceptibility and also the gapless spin excitation
spectrum observed by inelastic neutron scattering indicate 
the possibility of spin liquid ground state\cite{fujihala2020gapless,Liu_specific_heat}.

In this paper, we investigate its spin state by $^{27}$Al-NMR technique at low temperatures down to 0.6 K.  
Observed large NMR line width at low temperatures shows that the system has a magnetic 
ground state, which is consistent with previous reports\cite{fujihala2020gapless,Liu_specific_heat}.
However, the temperature dependence of $1/T_1$ obeys
the thermal activation law, suggesting the existence of a finite gap in the spin excitation spectrum.

\section{Experimental}\label{sec2}

Single phase polycrystalline ${\rm KCu_6AlBiO_4(SO_4)_5Cl}$ was synthesized
by the solid-state reaction described in ref. \cite{fujihala2020gapless}.
$^{27}$Al-NMR ($I=5/2$) experiment was performed on 16T superconducting magnet with the $^3$He cryocooler\cite{Matsui}.
The spectra were obtained by plotting the spin-echo amplitude against the applied magnetic field, which was varied slowly.
The longitudinal nuclear spin relaxation time $T_1$ was determined from the nuclear spin magnetization recovery curve, which
was obtained by plotting the nuclear spin magnetization against the iteration interval $\tau$,
and was fitted with the stretched exponential function ${1-e^{-(\tau/T_1)^\beta}}$, 
where $\beta$ is the temperature dependent constant index related to the inhomogeneity within the system\cite{Urano_vortex,Suzuki_muSR}.
$\beta$ takes value unity when the system is ideally pure, and decreases with increasing amount disorder.
Usually, the recovery curves of $I = 5/2$ nuclear spins obey the 3-exponential function or stretched function. 
However, for the present case, the single stretched exponential function well fitted to the observed data, so that
we adopted it.  Due to the appreciable amount of disorder in the system, as can been seen in the rather low value of $\beta$ 
may cause the distribution of $1/T_1$, and hence the smear-out of 3-exponential function into the single stretched exponential function.

The nuclear site of Al is located in the midst of two adjacent Kagom\'{e} planes facing each other, and is expected to probe the 
averaged magnetism of the nearest and the second nearest Cu's including
two inequivalent sites. 
This leads to the moderate value of the hyperfine coupling constant
$3A_{\rm an}=0.75({\rm T}/\mu_{\rm B})$, as will be described below.

\begin{figure}[h]%
\centering
\includegraphics[width=6cm,bb=110 70 510 470,clip]{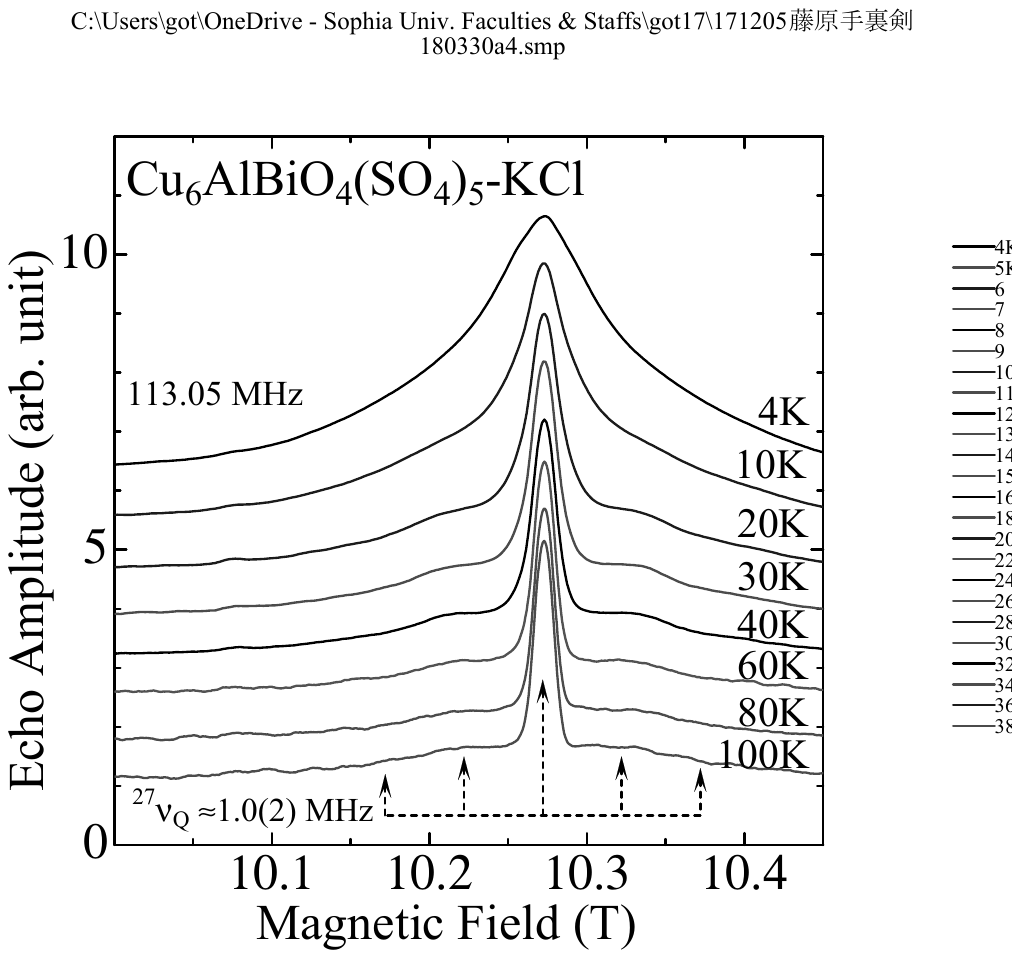}
\caption{Typical $^27$Al NMR spectra at various temperatures.  
Dashed arrows show the $90^{\circ}$-singular points 
in eqq-powder pattern of $I=5/2$ nucleus.}\label{fig2}
\end{figure}

\section{Results and Discussion}\label{sec3}
Figure \ref{fig2} shows typical profile of $^{27}$Al-NMR spectra at various temperatures.
The spectra were nearly symmetric and consisted with a sharp peak at the center and broad tails at both sides, 
which is a typical feature for the nucleus weakly perturbed with the electric quadrupole (eqq) 
interaction including some disorder\cite{Matsui}.
The interaction parameter $\nu_{\rm Q} = $ 1.0(2)MHz was deduced from the distance between small anomalies in the spectra,
corresponding to the satellite transitions of $I=5/2$ nuclear spin. 

The line width increased monotonically with lowering temperatures, while the position of the central peak
does not change at all.
This indicates that the isotropic part of the hyperfine coupling constant $A_{\rm iso}$ is zero within an experimental accuracy, 
and that the anisotropic part $A_{\rm an}$, mediated by the classical dipole-dipole interaction is dominant.

\begin{figure}[ht]%
\centering
\includegraphics[width=7cm,bb=30 380 530 780,clip]{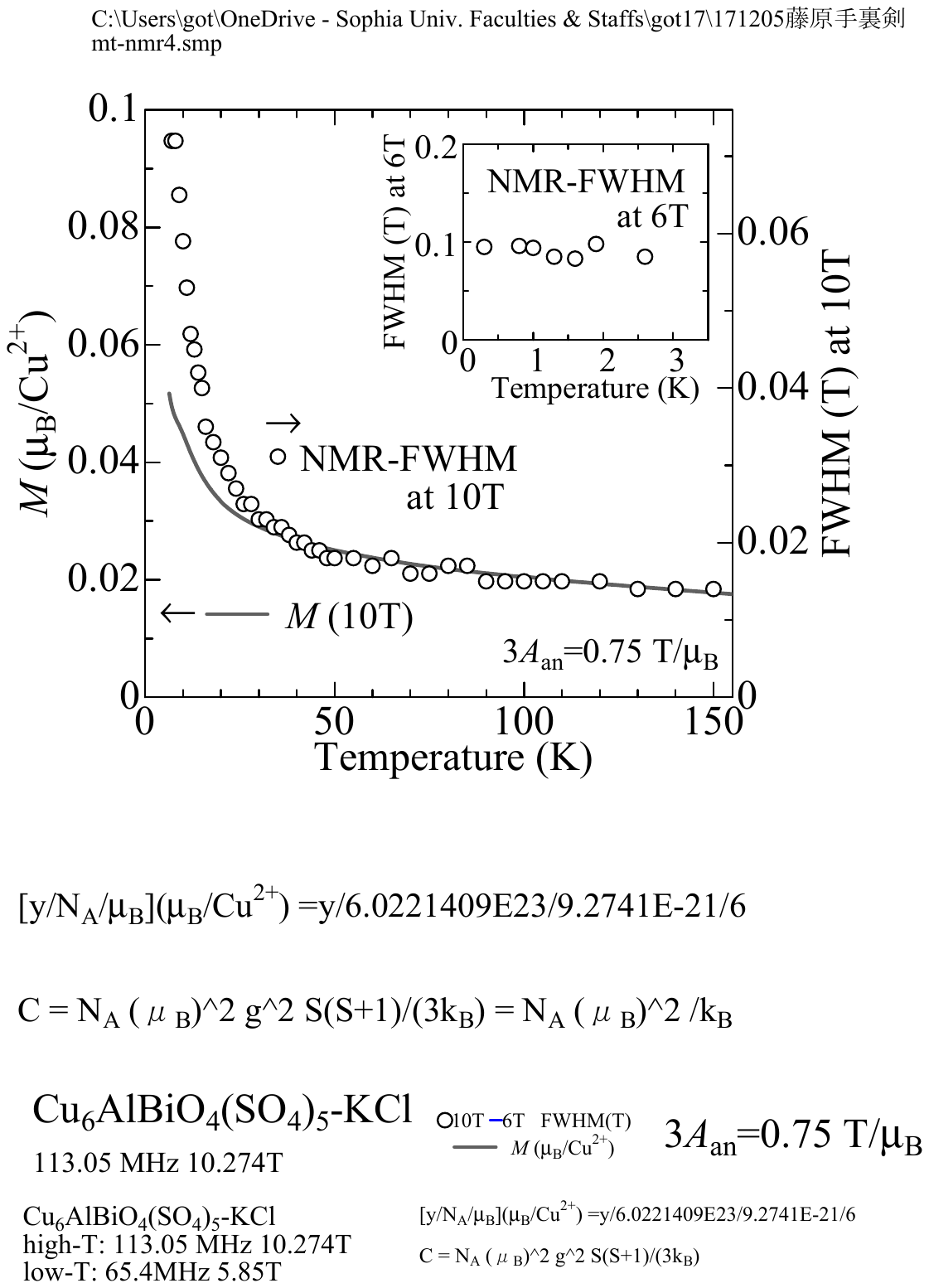}
\caption{Temperature dependence of the NMR line width (FWHM) under 10 T,
scaled with the uniform magnetization, estimated as $M = \chi B_{\rm NMR}$, 
where $\chi$ is the uniform susceptibility\cite{fujihala2020gapless}, and
$B_{\rm NMR}= 10 T$ is the measurement field of NMR line width.
(inset) NMR line width under 6 T down to 0.35 K.}\label{fig3}
\end{figure}

The increase in the line width is due to the inhomogeneous broadening, and 
is apparently magnetic origin. 
The size of $A_{\rm an} = 0.25 ({\rm T}/\mu_{\rm B})$
was determined by scaling the temperature dependence of uniform susceptibility with
that of line width defined as FWHM\cite{Inoue}.  
Their temperature dependences are shown in Fig. \ref{fig3}, where
the good scaling holds only in the high temperature region above 30 K.

The obtained value of $A_{\rm an}$ is consistent with the fact that the Al nucleus interacts 
with tens of near Cu spins, locating at distances between 4 - 7 ${\rm \AA}$ from the Al site
by the classical dipole-dipole interaction.
No characteristic powder patterns with a singular point
due to the dipole interaction was seen in NMR spectra,
because the contribution from number of Cu sites with different positions may
interfere with each other and smeared out the pattern.

As lowering the temperature, the scaling between $\chi$ and NMR line width ceased at 30 K, 
below which the increase in the width 
overwhelmed that in $\chi$, indicating the emergence of antiferromagnetic spin correlation,
which only suppresses the former.
The width continued to increase monotonically and saturated below 3 K, without showing any anomaly.
This smooth increasing behavior assures the absence of phase transition, the fact of which is consistent
with the previous reports\cite{fujihala2020gapless,Liu_specific_heat}.
The observed large value of NMR width at low temperatures indicates that
the ground state, even though paramagnetic, bears a large internal magnetic field under the
applied field of 10 T.
A rough estimation based on the obtained hyperfine coupling constant, the line width of 0.1 T
corresponds to the magnetic moment of $\mu = $0.2(1) $\mu_{\rm B}$.
This value is much larger than the reported value of uniform magnetization 0.03$\mu_{\rm B}$
measured under the magnetic field of 10 T\cite{fujihala2020gapless}, implying
that the appreciably large staggered moment is induced within the system.

\begin{figure}[h]%
\centering
\includegraphics[width=6cm,bb=60 340 460 740,clip]{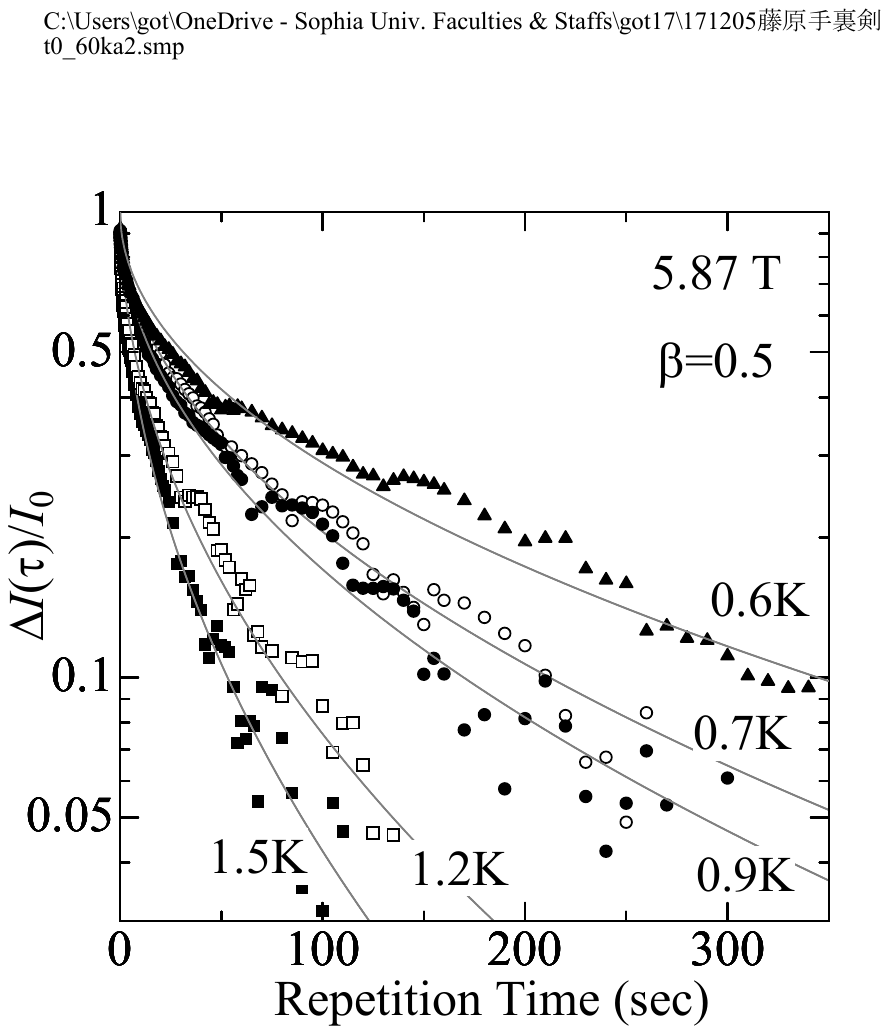}
\caption{Typical recovery curves in NMR-$T_1$ measurements. Solid curves show the theoretical recovery function with the stretched exponential function 
with the index $\beta = 0.5$.}\label{fig4}
\end{figure}

\begin{figure}[h]%
\centering
\includegraphics[width=7cm,bb=40 310 480 780,clip]{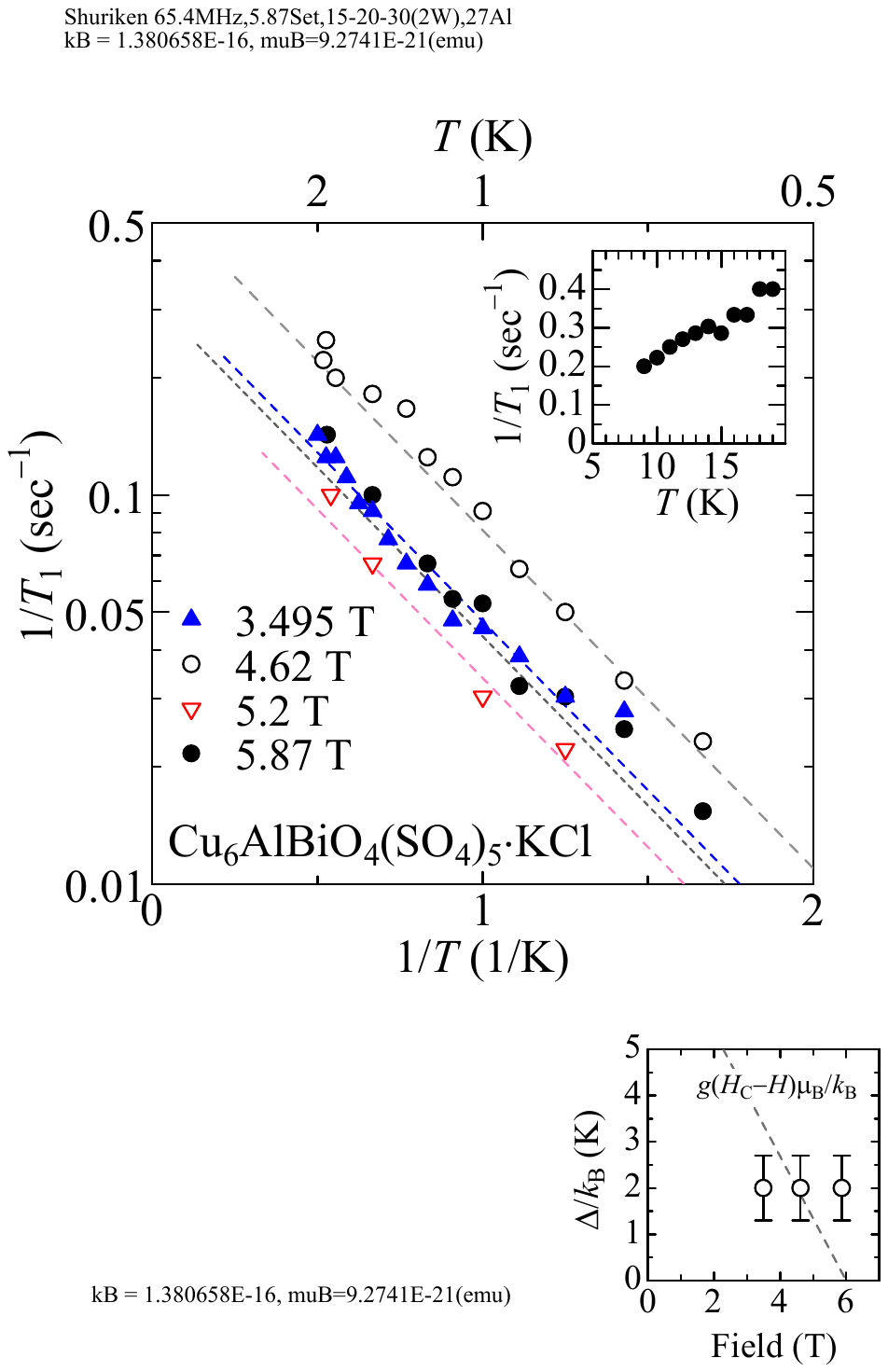}
\caption{Temperature dependence of 1/$T_1$ under various magnetic fields.
The inset shows high temperature behavior.  Dashed lines show thermal activation-type temeprature dependence with the gap $\Delta \simeq 2 K$.  }\label{fig5}
\end{figure}

Next, we investigated the dynamical spin correlation by means of NMR-$1/T_1$, which is the measure of 
the Larmor-frequency-Fourier component of spin fluctuation power spectrum.
Figure \ref{fig4} shows typical recovery curves of nuclear spin magnetization, which were 
analyzed by a single stretched exponential function with $\beta = 0.5$\cite{Urano_vortex}.
The temperature dependence of $1/T_1$ under various magnetic fields is shown in Fig. \ref{fig5}.
As lowering temperature, $1/T_1$ steeply decreases and obeys the thermal activation law below 2 K,
down to 0.35 K.
The value of activation gap is obtained from fitting to be $\Delta =$2.0(1) K, which is 
independent of the field between 3.495 and 5.87 T.

This observation clearly indicates that the present system has a gapped ground state.
However, its spin state must be different from the other gapped dimer systems such as
the valence bond crystal (VBC)\cite{matan2010pinwheel,ruegg2003bose}.
This is simply because the present system bears a large staggered moment at the lowest temperature, while
in the VBC, the system must become non magnetic by the formation of singlet dimers.
The saturating behavior of NMR line width below 3 K, as shown in the inset of Fig. 3, supports this idea.
That is, if in the VBC case, the line width is expected to be reduced rather than saturating at the large value.

Furthermore, the field-independent gap in the present system shows that the first excited state is 
not simply the triplet state of $S_z = 1$.
If that is the case, the gap must be reduced by the applied magnetic field, as is often observed in magnon BEC systems\cite{ruegg2003bose}.
We refer that the nematic state, for example, the triplet state of $S_z =0$, the Zeeman energy of which is 
field independent, may be one of the candidate\cite{nematic}.

At this stage, we cannot explain the mechanism of gap formation in this system, and it seems that more investigation 
on both experiments and theories must be necessary.  
Finally, we refer the dimer coverage on the present system. 
Unlike ordinary kagom\'{e} systems,
the square kagom\'{e} is non-bipartite, and hence it cannot be covered by dimers.
However, Siddharthan $et al.$ had pointed out that the system can be covered by dimers, if only the translational symmetry is broken
and the unit cell is elongated\cite{siddharthan2001_translational}.

\section{Conclusion}\label{sec4}
In summary, we have investigated $^{27}$Al-NMR on $S=1/2$ antiferromagnet KCu$_6$AlBiO$_4$(SO$_4$)$_5$Cl 
on the square-Kagom\'{e} lattice.  
From the detailed temperature dependence of the line width and $1/T_1T$, it was concluded that the system  
has a magnetic ground state with a finite energy gap $\Delta =$2.0(1) K, which is independent of the applied field.



\bmhead{Author Contributions} T.G. wrote the main manuscript text and performed NMR experiments.  M.F and S.M. prepared the sample.
All authors reviewed the manuscript.

\bmhead{Funding} This work was supported by Grant-in-Aid for Scientific Research (Grants No. 21H01955, 21K03452 and 19H00658) from
MEXT, Japan.

\bmhead{Availability of data and materials} The data that support the findings of this study are available from the
corresponding author upon reasonable request.

\section{Declaration}\label{sec4}

\bmhead{Ethics standard} Not applicable.

\bmhead{Competing interests} The authors declare no competing interests.

\bibliography{hf2023g}

\end{document}